\documentclass[aps,prd,twocolumn,showpacs,superscriptaddress,amsmath,longbibliography, nofootinbib]{revtex4-1}
\usepackage[dvips]{graphicx}
\usepackage{epstopdf}
\usepackage{xspace}

\usepackage{hyperref} % add hypertext capabilities 
\usepackage[usenames]{color} % color for proofread
\usepackage{ulem}
\usepackage{multirow}
\usepackage{enumitem}

\RequirePackage{lineno}

\graphicspath{{figures/}}

\begin{document}

%\pagewiselinenumbers
%\linenumbers

%\renewcommand{\thesection}{\Roman{section}}

%\title{{\Large NOT FOR DISTRIBUTION -- ONLY FOR INTERNAL USE}\\
%\vspace{0.2cm}
\title{
Search for Proton Decay via $p \rightarrow e^{+} \pi^{0}$ and $p \rightarrow \mu^{+} \pi^{0}$ in 0.31 megaton$\cdot$years exposure of the Super-Kamiokande Water Cherenkov Detector}

%\input{authors}
%%%% Authors
\newcommand{\AFFicrr}{\affiliation{Kamioka Observatory, Institute for Cosmic Ray Research, University of Tokyo, Kamioka, Gifu 506-1205, Japan}}
\newcommand{\AFFkashiwa}{\affiliation{Research Center for Cosmic Neutrinos, Institute for Cosmic Ray Research, University of Tokyo, Kashiwa, Chiba 277-8582, Japan}}
\newcommand{\AFFipmu}{\affiliation{Kavli Institute for the Physics and
Mathematics of the Universe (WPI), The University of Tokyo Institutes for Advanced Study,
University of Tokyo, Kashiwa, Chiba 277-8583, Japan }}
\newcommand{\AFFmad}{\affiliation{Department of Theoretical Physics, University Autonoma Madrid, 28049 Madrid, Spain}}
\newcommand{\AFFubc}{\affiliation{Department of Physics and Astronomy, University of British Columbia, Vancouver, BC, V6T1Z4, Canada}}
\newcommand{\AFFbu}{\affiliation{Department of Physics, Boston University, Boston, MA 02215, USA}}
\newcommand{\AFFbnl}{\affiliation{Physics Department, Brookhaven National Laboratory, Upton, NY 11973, USA}}
\newcommand{\AFFuci}{\affiliation{Department of Physics and Astronomy, University of California, Irvine, Irvine, CA 92697-4575, USA }}
\newcommand{\AFFcsu}{\affiliation{Department of Physics, California State University, Dominguez Hills, Carson, CA 90747, USA}}
\newcommand{\AFFcnm}{\affiliation{Department of Physics, Chonnam National University, Kwangju 500-757, Korea}}
\newcommand{\AFFduke}{\affiliation{Department of Physics, Duke University, Durham NC 27708, USA}}
\newcommand{\AFFfukuoka}{\affiliation{Junior College, Fukuoka Institute of Technology, Fukuoka, Fukuoka 811-0295, Japan}}
\newcommand{\AFFgifu}{\affiliation{Department of Physics, Gifu University, Gifu, Gifu 501-1193, Japan}}
\newcommand{\AFFgist}{\affiliation{GIST College, Gwangju Institute of Science and Technology, Gwangju 500-712, Korea}}
\newcommand{\AFFuh}{\affiliation{Department of Physics and Astronomy, University of Hawaii, Honolulu, HI 96822, USA}}
\newcommand{\AFFkek}{\affiliation{High Energy Accelerator Research Organization (KEK), Tsukuba, Ibaraki 305-0801, Japan }}
\newcommand{\AFFkobe}{\affiliation{Department of Physics, Kobe University, Kobe, Hyogo 657-8501, Japan}}
\newcommand{\AFFkyoto}{\affiliation{Department of Physics, Kyoto University, Kyoto, Kyoto 606-8502, Japan}}
\newcommand{\AFFmiyagi}{\affiliation{Department of Physics, Miyagi University of Education, Sendai, Miyagi 980-0845, Japan}}
\newcommand{\AFFnagoya}{\affiliation{Solar Terrestrial Environment Laboratory, Nagoya University, Nagoya, Aichi 464-8602, Japan}}
\newcommand{\AFFpol}{\affiliation{National Centre For Nuclear Research, 00-681 Warsaw, Poland}}
\newcommand{\AFFsuny}{\affiliation{Department of Physics and Astronomy, State University of New York at Stony Brook, NY 11794-3800, USA}}
\newcommand{\AFFokayama}{\affiliation{Department of Physics, Okayama University, Okayama, Okayama 700-8530, Japan }}
\newcommand{\AFFosaka}{\affiliation{Department of Physics, Osaka University, Toyonaka, Osaka 560-0043, Japan}}
\newcommand{\AFFregina}{\affiliation{Department of Physics, University of Regina, 3737 Wascana Parkway, Regina, SK, S4SOA2, Canada}}
\newcommand{\AFFseoul}{\affiliation{Department of Physics, Seoul National University, Seoul 151-742, Korea}}
\newcommand{\AFFshizuokasc}{\affiliation{Department of Informatics in
Social Welfare, Shizuoka University of Welfare, Yaizu, Shizuoka, 425-8611, Japan}}
\newcommand{\AFFskk}{\affiliation{Department of Physics, Sungkyunkwan University, Suwon 440-746, Korea}}
\newcommand{\AFFtokyo}{\affiliation{The University of Tokyo, Bunkyo, Tokyo 113-0033, Japan }}
\newcommand{\AFFtodai}{\affiliation{Department of Physics, University of Tokyo, Bunkyo, Tokyo 113-0033, Japan }}
\newcommand{\AFFtoronto}{\affiliation{Department of Physics, University of Toronto, 60 St., Toronto, Ontario, M5S1A7, Canada }}
\newcommand{\AFFtriumf}{\affiliation{TRIUMF, 4004 Wesbrook Mall, Vancouver, BC, V6T2A3, Canada }}
\newcommand{\AFFtokai}{\affiliation{Department of Physics, Tokai University, Hiratsuka, Kanagawa 259-1292, Japan}}
\newcommand{\AFFtsinghua}{\affiliation{Department of Engineering Physics, Tsinghua University, Beijing, 100084, China}}
\newcommand{\AFFuw}{\affiliation{Department of Physics, University of Washington, Seattle, WA 98195-1560, USA}}

\AFFicrr
\AFFkashiwa
\AFFmad
\AFFbu
\AFFubc
\AFFbnl
\AFFuci
\AFFcsu
\AFFcnm
\AFFduke
\AFFfukuoka
\AFFgifu
\AFFgist
\AFFuh
\AFFkek
\AFFkobe
\AFFkyoto
\AFFmiyagi
\AFFnagoya
\AFFpol
\AFFsuny
\AFFokayama
\AFFosaka
\AFFregina
\AFFseoul
\AFFshizuokasc
\AFFskk
\AFFtokai
\AFFtokyo
\AFFtodai
\AFFipmu
\AFFtoronto
\AFFtriumf
\AFFtsinghua
\AFFuw

%%%%%%%%%%%%%%%%%%%%%%%%%%%%%%%%%%%%%%%%%%%%%%%%%%%%%%%%%%%%%%%%%%%%
%ICRR
\author{K.~Abe}
\AFFicrr
\AFFipmu
\author{Y.~Haga}
\AFFicrr
\author{Y.~Hayato}
\AFFicrr
\AFFipmu
\author{M.~Ikeda}
\AFFicrr
\author{K.~Iyogi}
\AFFicrr 
\author{J.~Kameda}
\author{Y.~Kishimoto}
\author{M.~Miura} 
\author{S.~Moriyama} 
\author{M.~Nakahata}
\AFFicrr
\AFFipmu 
% ...

\collaboration{The Super-Kamiokande Collaboration}
\noaffiliation
%%%%

\date{\today}

\begin{abstract}

We have searched for proton decay via $p \rightarrow e^{+} \pi^{0}$ and $p
\rightarrow \mu^{+} \pi^{0}$ using Super-Kamiokande data from April 1996 to
March 2015, 0.306 megaton$\cdot$years exposure in total.  The atmospheric
neutrino background rate in Super-Kamiokande IV is reduced to almost half
that of phase I-III by tagging neutrons associated with neutrino
interactions.  The reach of the proton lifetime is further enhanced by
introducing new signal criteria that select the decay of a proton in a
hydrogen atom.  No candidates were seen in the $p \rightarrow e^{+} \pi^{0}$
search. Two candidates that passed all of the selection criteria for $p
\rightarrow \mu^{+} \pi^{0}$ have been observed, but these are consistent
with the expected number of background events of 0.87.  Lower limits on the
proton lifetime are set at $\tau/B(p \rightarrow e^{+} \pi^{0}) > 1.6 \times
10^{34}$ years and $\tau/B(p \rightarrow \mu^{+} \pi^{0}) > 7.7 \times
10^{33}$ years at 90\% confidence level.

\end{abstract}

\maketitle
%\tableofcontents
\newpage

\section{INTRODUCTION}

Grand Unified Theories (GUTs)~\cite{Pati:1973uk,*Pati:1973rp,*Georgi:1974sy}
are motivated by the apparent convergence of the running couplings of the
strong, weak, and electromagnetic forces at high energy
($10^{15}-10^{16}$\,GeV). Such a high energy scale is out of the reach of
accelerators; however, a general feature of GUTs is their prediction of the
instability of protons by baryon number violating decays.  The grand
unification idea is successful in many aspects; these include an
understanding of electric charge quantization, the co-existence of quarks and
leptons and their quantum numbers, and as an explanation of the scale of the
neutrino masses.  Proton decay now remains as a key missing piece of evidence
for grand unification~\cite{Babu:2013jba}.

In GUTs, nucleon decay can proceed via exchange of a massive gauge boson
between two quarks. The favored gauge-mediated decay mode in many GUTs is $p
\rightarrow e^{+} \pi^{0}$.  On the other hand, the flipped $SU(5)$ GUT
model~\cite{Ellis:2002vk} predicts that the $p \rightarrow \mu^{+} \pi^{0}$
mode can have a branching ratio comparable to that of the $p \rightarrow
e^{+} \pi^{0}$ mode.  Water Cherenkov detectors are suitable for these decay
modes because all final state particles after the proton decay are detectable
since they are above the Cherenkov threshold, enabling reconstruction of the
proton mass and momentum to distinguish these events from atmospheric
neutrino backgrounds.  The dominant inefficiency comes from Fermi momentum of
protons and pion interactions inside the nucleus, which distorts the
reconstructed proton mass and momentum distributions. However, the two
hydrogen atoms in a water molecule are outside of the oxygen nucleus; these
act as free protons which are not subject to the nuclear effects.  As a
result, water Cherenkov detectors can achieve high efficiency for $p
\rightarrow e^{+} \pi^{0}$ and $p \rightarrow \mu^{+} \pi^{0}$ searches.

In the minimal SU(5) GUT~\cite{Langacker:1994vf}, the predicted proton
lifetime to $e^+\pi^0$ is $10^{31\pm 1}$ years, which has been ruled out by
experimental results from IMB~\cite{McGrew:1999nd},
Kamiokande~\cite{Hirata:1989kn}, and
Super-Kamiokande~\cite{Shiozawa:1998si,Nishino:2009aa,Nishino:2012bnw}.
However, longer lifetimes for this decay mode ($\sim 10^{35}$ years) are
predicted by other classes of GUTs {\it{e.g.}}, minimal SUSY
SU(5)~\cite{Hisano:1992jj,*Murayama:2001ur}, flipped
SU(5)~\cite{Ellis:2002vk}, SO(10)~\cite{Babu:2015bna,*Altarelli:2013aqa},
{\it{etc.}}, which are subject to experimental tests.  The experimental
searches for the gauge-mediated proton decays are further motivated by the
discovery of a Higgs-like boson with a mass around
125\,GeV/c$^{2}$~\cite{Aad:2015zhl,Hisano:2013exa,*Fukuda:2015pra}.  This paper
describes the search for $p \rightarrow e^{+} \pi^{0}$ and $p \rightarrow
\mu^{+} \pi^{0}$ by improved analysis techniques with 0.306
megaton$\cdot$years of Super-Kamiokande data.

\section{SUPER-KAMIOKANDE DETECTOR}
Super-Kamiokande (SK) is a large upright cylindrical water Cherenkov
detector, 39\,m in diameter and 41\,m in height, containing 50\,ktons of pure
water. Details of the detector have been described in
Ref.~\cite{Fukuda:2002uc,*Abe:2013gga}. The previous publication of $p
\rightarrow e^{+} \pi^{0}$ and $p \rightarrow \mu^{+} \pi^{0}$
limits~\cite{Nishino:2012bnw} reported results using 220 kiloton$\cdot$years
exposure with 90\% confidence level lower limits on the proton lifetime set
at 1.3$\times10^{34}$ and 1.1$\times 10^{34}$ years, respectively.  In the
SK-I period, the photocathode coverage inside the inner detector was 40\%;
this was reduced to 19\% during SK-II. After production and installation of
replacement 50-cm photomultiplier tubes (PMTs), the photocathode coverage
inside the detector was recovered to the original 40\% in 2006. The period
between July 2006 and September 2008 is defined as SK-III. In the summer of
2008, the detector readout electronics were upgraded with improved
performance, including a ``deadtime free'' data acquisition system that
records all successive PMT hit
information~\cite{Nishino:2009zu,*Yamada:2010zzc}.  This has been the
configuration of the detector since September 2008; it is called SK-IV.  The
new configuration of the detector contributes to improved tagging efficiency
of Michel decay electrons in SK-IV.  The tagging efficiency of Michel
electrons is estimated to be 99\% for SK-IV and 82\% for the period before SK-IV
by using $p \rightarrow \mu^{+} \pi^{0}$ MC samples.  It also enables tagging
of neutrons in SK-IV.  The signature of the neutron, 2.2\,MeV gamma ray
emission from the neutron capture on hydrogen with a mean capture time of
200\,$\mu$sec, is too faint to be triggered by the data acquisition system
used in SK-I through SK-III.  The upgraded electronics in SK-IV adopt a
trigger-less readout scheme to record every hit, including all hits by
2.2\,MeV gamma rays.  A software trigger is then issued after every
fully-contained event to save all hits within a 500\,$\mu$sec timing window
for physics analyses.  In this paper, we use data from April 1996 up to March
2015, corresponding to 4973 live days or 306.3\,kton$\cdot$years in total by
summing up 91.7, 49.2, 31.9, 133.5\,kton$\cdot$years of SK-I, II, III, and IV
data.
%# SK-I    1489d   91.7ktyr
%# SK-II    799d   49.2ktyr
%# SK-III   518d   31.9ktyr
%# SK-IV   2167d  133.5ktyr 
%# total   4973d  306.3ktyr 

\section{NEUTRON TAGGING} 
Neutron tagging can benefit proton decay analyses, providing an additional
handle for rejecting the atmospheric neutrino interactions that are the main
background to proton decays.  Atmospheric neutrino interactions are often
accompanied by neutrons, while the probability of a neutron being emitted by
deexcitation of a nucleus after proton decay is rather small, and no neutron
is produced in the decay of a free proton.  Furthermore, these proton decay modes do not  produce neutrons in secondary interactions in water because the final state particles in these decay modes are lepton and gammas. The neutron tagging algorithm was
originally developed to identify anti-neutrino interactions, in which a
neutron is often emitted; the supernova relic neutrino
search~\cite{Zhang:2013tua}, the cosmic-ray-muon spallation background
measurement~\cite{Super-Kamiokande:2015xra}, and the neutrino oscillation
analysis~\cite{Irvine:2014hja} have been improved by this technique.  To find
2.2~MeV $\gamma$ candidates, we search for hit clusters with $\ge 7$ hits
within a 10~ns sliding time window, after time-of-flight (TOF) subtraction
using the vertex of the prompt neutrino interaction.  Sixteen variables
described in Ref.~\cite{Irvine:2014hja} are input to a neural network to
distinguish the 2.2~MeV $\gamma$ signal from background.  The efficiency of
neutron tagging by this method is estimated to be 20.5$\pm$2.1\% with a
mis-tagging probability of 1.8\%.  The performance of the neutron tagging was
confirmed by introducing a neutron source (Americium-Beryllium) into the SK
tank~\cite{Zhang:2013tua}.

\section{SIMULATION} 
Our simulation of the atmospheric neutrino cross sections and flux is modeled
by NEUT~\cite{Hayato:2002sd,*Mitsuka:2007zz,*Mitsuka:2008zza} and
HKKM~\cite{Honda:2011nf}.  In the case of atmospheric neutrino interactions,
neutrons are generated in: primary interactions (10\%), hadron and meson
interactions in the nucleus (17\%), and interactions of hadrons in water
(73\%).  For hadron-water interactions at energies below 3.5 GeV, our
simulation uses the Nucleon-Meson-Transport-Code (NMTC), which is based on
the Bertini intra-nuclear hadronic cascade model~\cite{Bertini:1970zs}.  For
low energy ($<$ 20 MeV) neutron propagation, the Monte Carlo Ionization
Chamber Analysis Package (MICAP) code~\cite{Micap} is employed.  Thermalized
neutrons are then simulated until capture on hydrogen and emission of 2.2 MeV
$\gamma$-rays.  Neutrons can also be generated via the deexcitation process
and muon capture in water. These are not taken into account in our
atmospheric neutrino simulation, but are instead included in our uncertainty
by comparison of data and atmospheric $\nu$ Monte Carlo.  In order to include
all possible low energy backgrounds, real data taken by a random trigger are
added into the simulated neutron data.

The proton decay simulation has been updated since the previous
paper~\cite{Abe:2014mwa} by implementation of proton and neutron emission at
deexcitation from s-state~ \cite{Ejiri:1993rh}, and 7.5\% of $p \rightarrow
e^{+} \pi^{0}$ events in oxygen are accompanied by a neutron.

\begin{figure*}[htbp]
 \begin{center}
   \includegraphics[width=90mm]{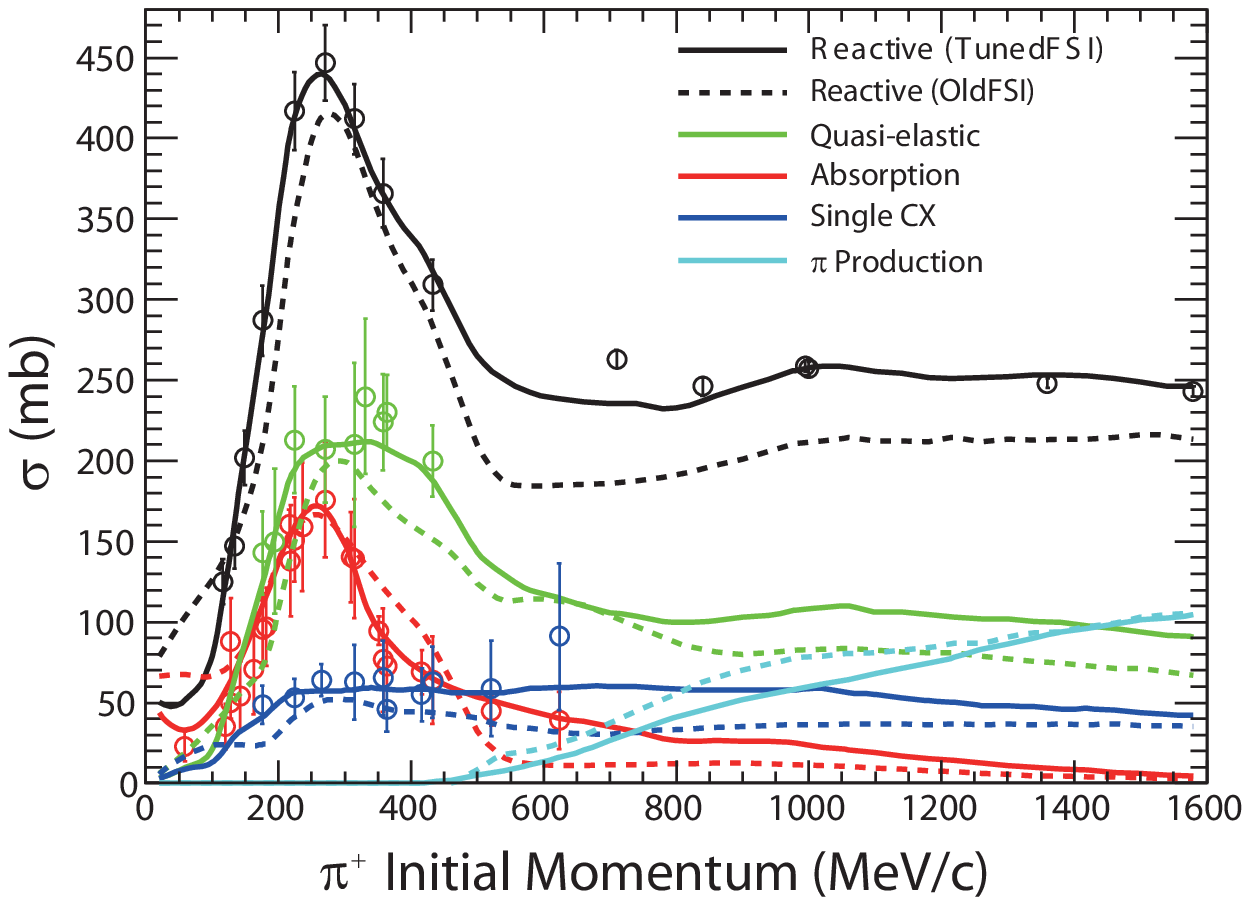}
   \includegraphics[width=85mm]{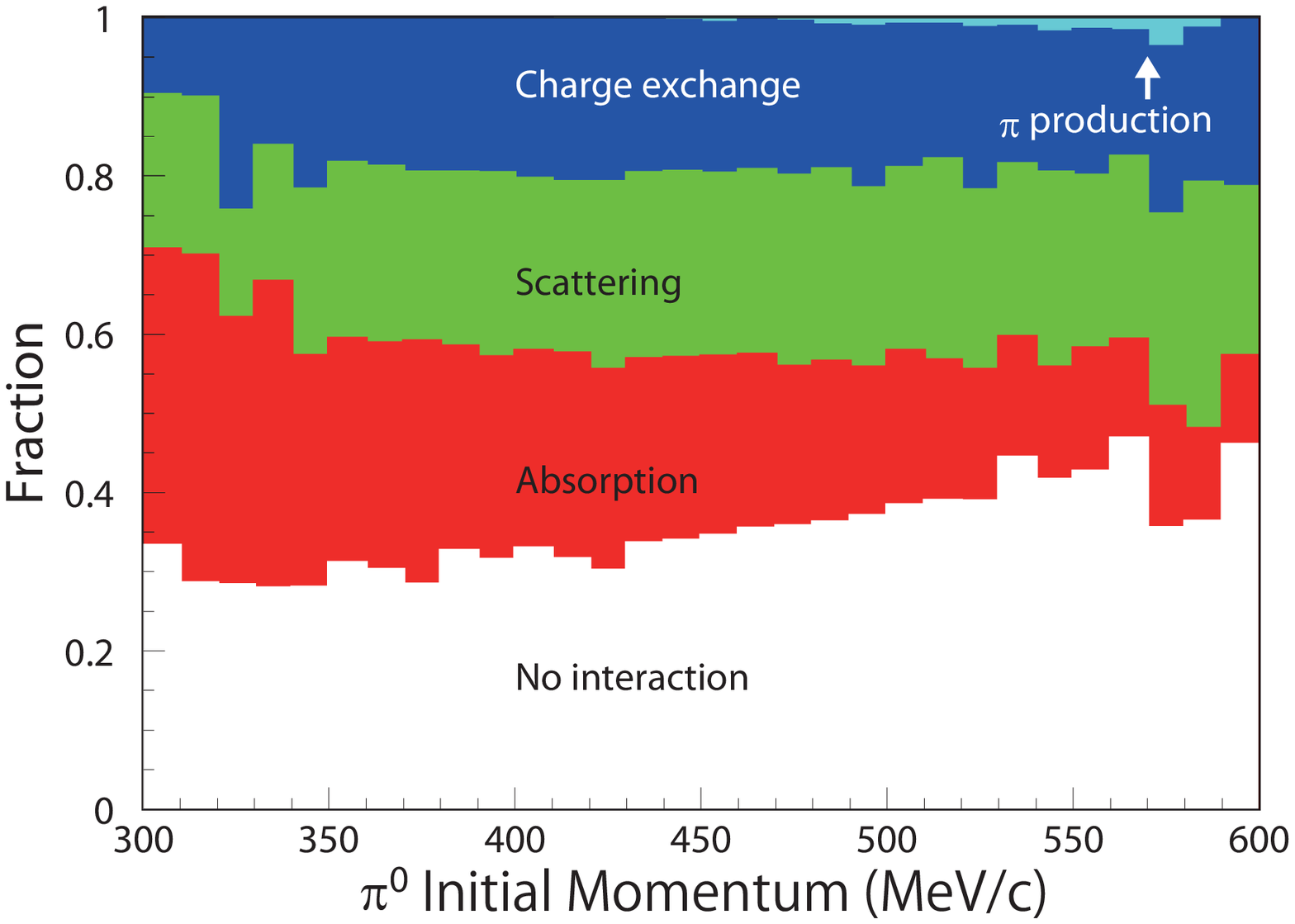}
 \end{center}
 \caption{ \protect \small (color online) The left panel shows $\pi^{+}$
   cross sections on $^{12}$C as a function of momentum.  Open circles
   indicate existing data~\cite{Giannelli:2000zy, *Saunders:1996ic,
     *Navon:1983xj, *Bowles:1981my, *Ashery:1981tq, *Ashery:1984ne,
     *Ashery:1984ne, *Levenson:1983xu, *Jones:1993ps, *Chavanon:1969zu,
     *Takahashi:1995hs, *Allardyce:1973ce, *Cronin:1957zz, *Aoki:2007zzb,
     *Rahav:1991vi}, solid and dashed lines show after and before the tune,
   respectively. The colors correspond to the total cross section and
   exclusive channels: total (black), elastic scattering (green), absorption
   (red), charge exchange (blue), and $\pi$ production (light blue). The
   right panel shows cumulative fractions of neutral pion final state
   interactions as a function of momentum generated by $p \rightarrow e^{+}
   \pi^{0}$ MC.  The fractions of events that undergo charge exchange,
   multiple $\pi$ production, scattering, and absorption are shown by the
   various shades as labeled. Neutral pions that exit the nucleus without
   experiencing any final state interactions are indicated by the portion
   labeled ``No interaction.''  }
  \label{pion}
\end{figure*}

A $\pi^0$ produced by bound proton decay interacts via the strong interaction
and there is a significant probability of re-interaction within the nucleus
prior to escape. NEUT also simulates the pion final state interactions
($\pi$-FSI).  The NEUT $\pi$-FSI model is a microscopic cascade where the
pion is propagated classically through the nuclear medium in finite steps; it
is capable of simulating various nuclei.  The $\pi$-FSI model has been tuned
to various $\pi^{\pm}$-nucleus experimental data including C, O, Al, Fe, and
has been updated~\cite{Abe:2013xua} since the previous
paper~\cite{Nishino:2012bnw}.  The left panel in Figure~\ref{pion} shows the
$\pi^{+}$ cross sections of the external data and MC on $^{12}$C
~\cite{Giannelli:2000zy, *Saunders:1996ic, *Navon:1983xj, *Bowles:1981my,
  *Ashery:1981tq, *Ashery:1984ne, *Ashery:1984ne, *Levenson:1983xu,
  *Jones:1993ps, *Chavanon:1969zu, *Takahashi:1995hs, *Allardyce:1973ce,
  *Cronin:1957zz, *Aoki:2007zzb, *Rahav:1991vi}, which has the largest number
of data points.  After tuning, the estimated rates of neutral pion absorption
and charge exchange have increased around 500 MeV/c, which affects the
distribution of the number of rings and causes a 5\% reduction of estimated
signal efficiency.  Large angle scattering in the nucleus has also increased,
which modifies the reconstructed proton mass and momentum distributions,
resulting in a 7\% reduction in estimated selection efficiency for $p
\rightarrow e^{+} \pi^{0}$ and $p \rightarrow \mu^{+} \pi^{0}$.  The right
panel of Figure~\ref{pion} shows the fraction of $\pi^0$ interactions as a
function of the pion momentum after the tune. The same plot for the previous
simulation can be found in~\cite{Abe:2013lua}.
    
\section{SELECTION CRITERIA} 
\begin{figure}[htbp]
 \begin{center}
   \includegraphics[width=90mm]{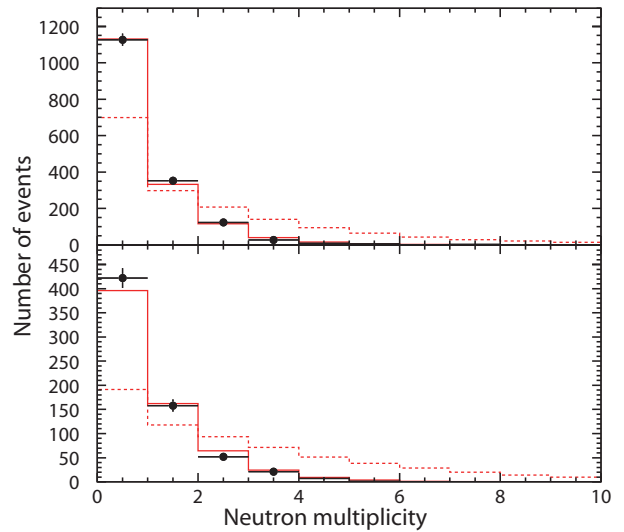}
 \end{center}
 \caption{ \protect \small Distribution of the number of tagged neutrons for
   $p \rightarrow e^{+} \pi^{0}$ (top) and $p \rightarrow \mu^{+} \pi^{0}$
   (bottom) after applying (Cut-1) through (Cut-5), excluding the signal
   region defined by (Cut-6). Dots show SK-IV data with 133.5
   kton$\cdot$years exposure, the solid histogram shows the multiplicity of
   tagged neutrons per event in the atmospheric $\nu$ MC, and the dashed
   histogram shows the true multiplicity of neutrons per event.  }
  \label{neutron}
\end{figure}

The following cuts are applied to signal MC, atmospheric $\nu$ MC, and data:
\begin{enumerate}[labelindent=10pt,leftmargin=!]
\item[(Cut-1)]{events must be fully contained in the inner detector with the
  vertex position within the fiducial volume, which is defined as 2 meters
  inward from the detector walls (FCFV),}
\item[(Cut-2)]{there must be 2 or 3 Cherenkov rings,}
\item[(Cut-3)]{all rings must be electron-like for $p \rightarrow e^{+}
  \pi^{0}$ and one ring must be $\mu$-like for $p \rightarrow \mu^{+}
  \pi^{0}$ ,}
\item[(Cut-4)]{there must be no Michel electrons for $p \rightarrow e^{+}
  \pi^{0}$ and one electron for $p \rightarrow \mu^{+} \pi^{0}$,}
\item[(Cut-5)]{reconstructed $\pi^0$ mass should be 85~$<
  M_{\pi^0}<$~185~MeV/$c^2$ for 3 ring events,}
\item[(Cut-6)]{reconstructed total mass should be $800<
  M_{tot}<1050$~MeV/$c^2$ and reconstructed total momentum $P_{tot}$ should
  be less than 250~MeV/{\it{c}},}
\end{enumerate}
and for SK-IV only,
\begin{enumerate}[labelindent=10pt,leftmargin=!]
\item[(Cut-7)]{there must be no tagged neutrons.}
\end{enumerate}
After (Cut-1), 41k FCFV events remain in data for the entire period. 
The events which contain two rings are allowed in (Cut-2) because one gamma ray from $\pi^{0}$ may fail to reconstruct if opening angle of the two gammas is small, or energy of the gamma is small. The fraction of 2 ring and 3 ring events in $p \rightarrow e^{+}
  \pi^{0}$ MC before (Cut-5) are 45\% and 55\%, respectively.  
The requirement (Cut-7) rejects about 50\% of the background but reduces
selection efficiency by 3\%, in which both true and fake neutrons contribute.  Figure~\ref{neutron} shows the number of tagged
neutrons for $p \rightarrow e^{+} \pi^{0}$ (upper) and $p \rightarrow \mu^{+}
\pi^{0}$ (lower) after applying (Cut-1) through (Cut-5), excluding the signal
region defined by (Cut-6).  These figures show that the neutron multiplicity
in the atmospheric neutrino MC (solid histogram) agrees well with data
(dots).  The dashed histograms show the true number of captured neutrons in
the MC, indicating that additional background rejection can be achieved if
the neutron tagging efficiency is improved in the future when gadolinium is
dissolved in the SK water~\cite{Beacom:2003nk}.

\begin{figure*}[htbp]
 \begin{center}
   \includegraphics[width=180mm]{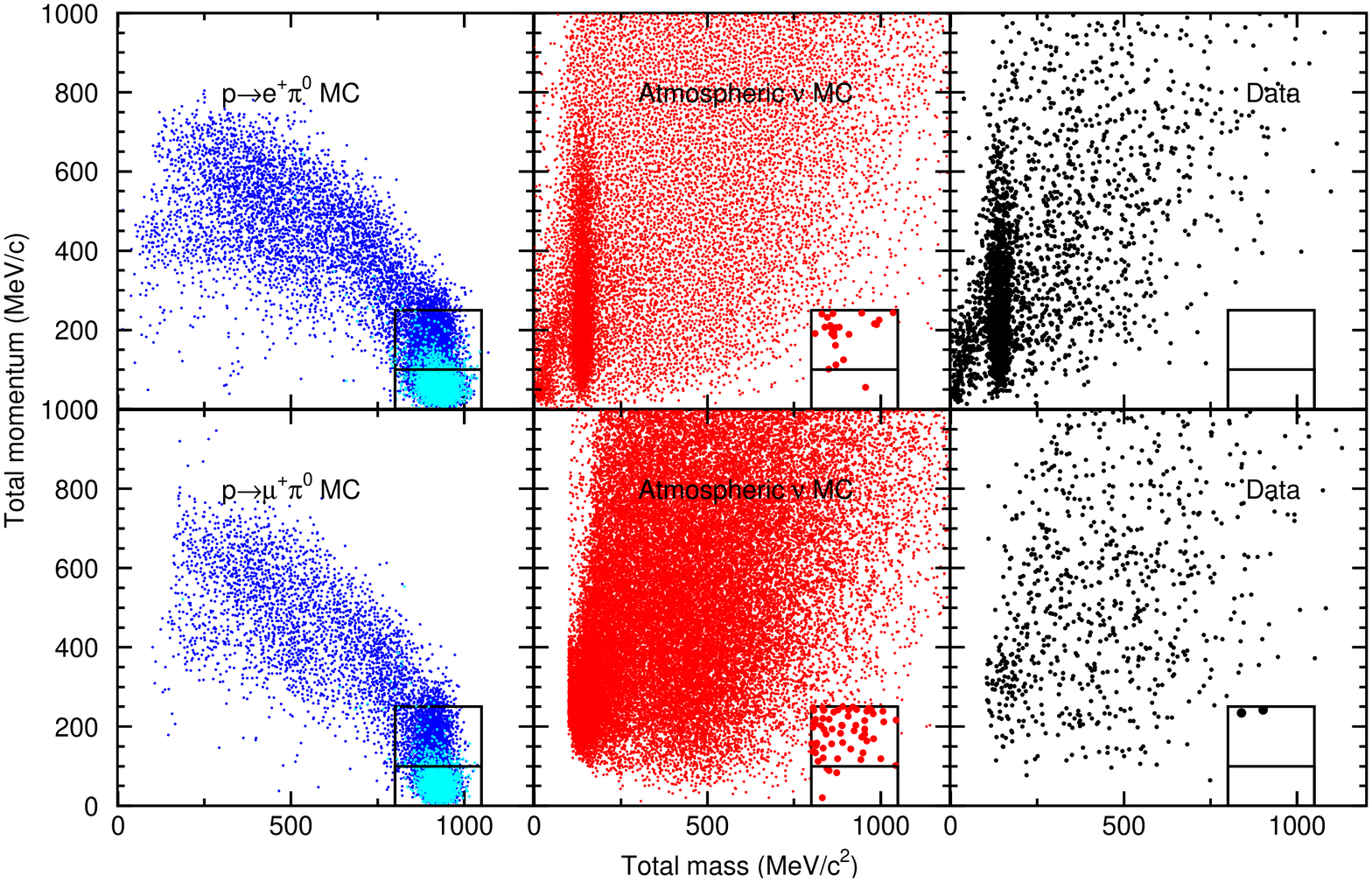}
 \end{center}
 \caption{ \protect \small (color online) Reconstructed proton mass
   {\it{vs.}} total momentum for $p \rightarrow e^{+} \pi^{0}$ (top) and $p
   \rightarrow \mu^{+} \pi^{0}$ (bottom) after all cuts except (Cut-6) .  The
   left panels show signal MC, where light blue corresponds to free protons
   and dark blue is bound protons. The middle panels show atmospheric $\nu$
   MC corresponding to 500 years live time of SK, and the right panels show
   SK-I to SK-IV data which contain 3408 and 1180 events for $p \rightarrow e^{+} \pi^{0}$  and $p \rightarrow \mu^{+} \pi^{0}$, respectively. The dot size is enlarged in the signal box.}
  \label{epi0-mupi0-scat}
\end{figure*}

\begin{figure*}[htbp]
 \begin{center}
   \includegraphics[width=180mm]{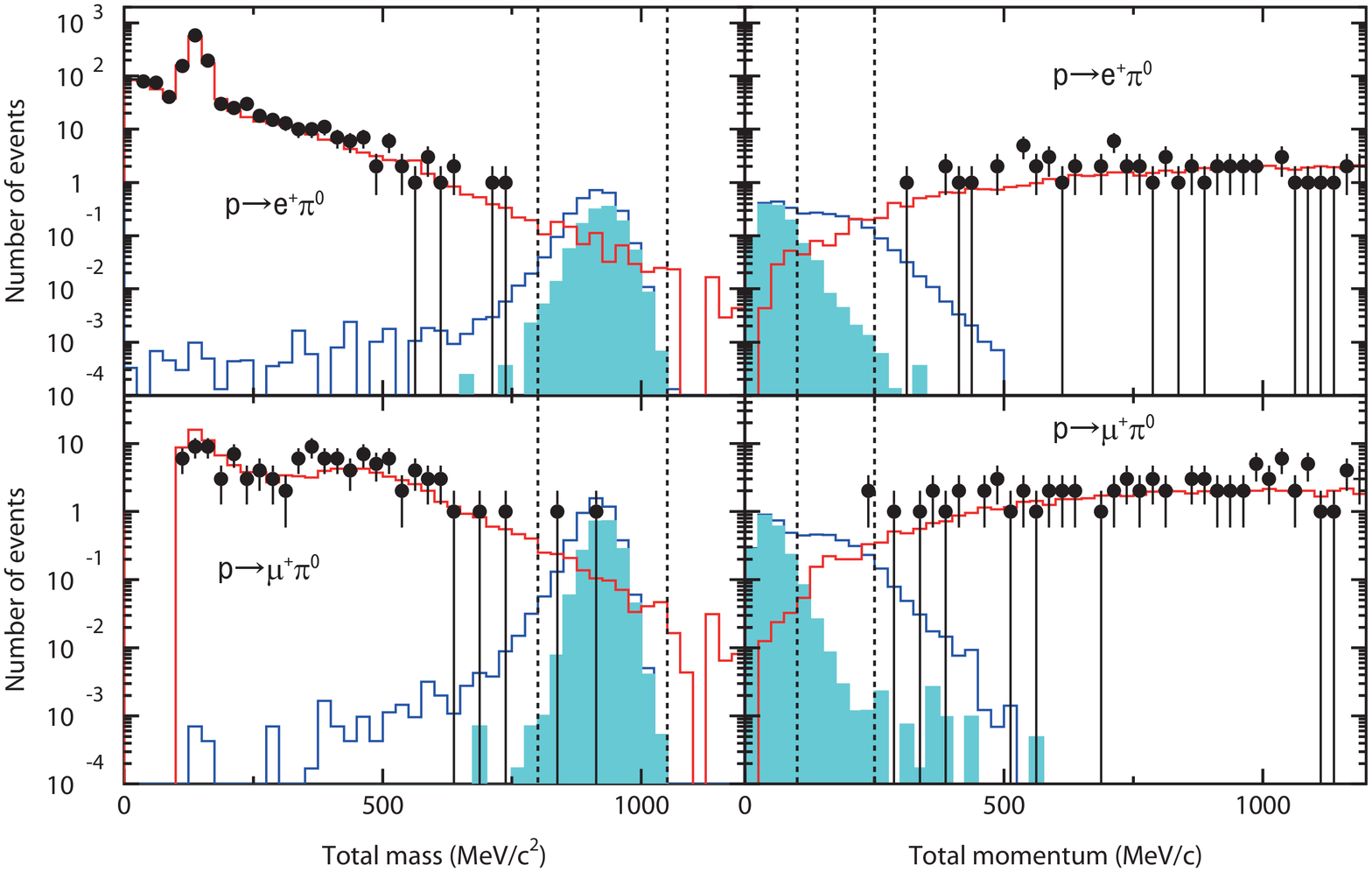}
 \end{center}
 \caption{ \protect \small (color online) Distributions of reconstructed
   invariant mass (left) and total momentum (right) for $p \rightarrow e^{+}
   \pi^{0}$ in the top panels and for $p \rightarrow \mu^{+} \pi^{0}$ in the
   bottom panels, after all selection cuts except cuts on the plotted
   variable. The dark blue histograms correspond to 90\% confidence level
   allowed signal and the histograms filled by light blue show the portion
   contributed by free proton decay. The red histograms show atmospheric
   $\nu$ MC, and the dots are data with 0.306 Mton$\cdot$years exposure.
   Vertical dashed lines indicate the signal regions.  The peak around 150
   MeV/$c^{2}$ in the total mass distribution of atmospheric $\nu$ and data
   in the left top panel arises from $\pi^{0}$ decays.  }
  \label{1d-both}
\end{figure*}

A new analysis technique~\cite{Nakamura:2000kz} is applied in this paper.
The signal region defined by (Cut-6) is divided into two regions:
$P_{tot}<100$~MeV/{\it{c}} and $100\le P_{tot} <250$~MeV/{\it{c}}. The region
below 100~MeV/{\it{c}} is dominated by free protons, and the region $100\le
P_{tot} <250$~MeV/{\it{c}} is dominated by bound protons.  A reduced
systematic error for $<100$~MeV/{\it{c}} can be achieved because the initial
protons and the products of the proton decay do not suffer from any of the
various nuclear interactions.  In addition, background contamination from
atmospheric neutrinos is concentrated in the $100\le
P_{tot}<250$~MeV/{\it{c}}, while the region below 100~MeV/{\it{c}} is nearly
background free.  Figure~\ref{epi0-mupi0-scat} shows the reconstructed proton
mass {\it{vs.}} total momentum distribution for signal and atmospheric $\nu$
MC, combining all data from SK-I through SK-IV.  In the signal MC plots, the
light blue dots show the free proton case and the dark blue dots show bound
protons.  Figure~\ref{1d-both} shows one-dimensional distributions of
reconstructed proton mass and momentum of $p \rightarrow e^{+} \pi^{0}$ and
$p \rightarrow \mu^{+} \pi^{0}$ for the signal, atmospheric $\nu$ MC, and
data after all cuts except the cut on the plotted variable.  The data and the
atmospheric $\nu$ MC agree in both cases. 

The new two-box analysis achieves
better discovery reach. For example, the 3$\sigma$ discovery reach in the proton lifetime for $p \rightarrow e^{+} \pi^{0}$ is 13.3\% higher than the conventional single-box
analysis for the current exposure, and 20.9\% higher for an exposure of 1
megaton$\cdot$year, which may be achieved by the next generation of
detectors~\cite{Abe:2011ts}.

\section{RESULTS} 
Table~\ref{result} shows efficiency, expected number of background events,
and number of observed events for each region of total momentum for $p
\rightarrow e^{+} \pi^{0}$ and $p \rightarrow \mu^{+} \pi^{0}$. The signal
efficiencies in SK-I and -II are decreased in comparison to the previous
paper due to the updated $\pi$-FSI model.  Efficiencies for $p \rightarrow
\mu^{+} \pi^{0}$ in SK-IV are significantly higher than the other data-taking
periods because of Michel electron tagging efficiency that was achieved with
the new electronics.  The background rate in SK-I/II/III is consistent
with the measurement in the 1 kton water Cherenkov detector using the K2K
accelerator neutrino beam~\cite{Mine:2008rt}. Background rates in SK-IV for
both decay modes are almost half that of SK-I/II/III, which is a result of
the contribution from neutron tagging.  In the entire signal region, with
306.3 kiloton$\cdot$years exposure, the expected background for $p
\rightarrow e^{+} \pi^{0}$ is 0.61 events (0.07 and 0.54 events in lower and
higher momentum box, respectively) and no candidates were observed, while for
$p \rightarrow \mu^{+}\pi^{0}$ the estimated background is 0.87 events (0.05
and 0.82 events in lower and higher momentum box, respectively) and two
candidates have been observed in the higher momentum box.  Assuming a Poisson
distribution with mean 0.87, the probability to see $\ge 2$ events is 22\%
and is still consistent with background.  The first candidate was observed at
257~kton$\cdot$years exposure and the second candidate was observed at
277~kton$\cdot$years exposure. A Kolmogorov-Smirnov test is carried out to
examine the assumption of a constant event rate. The obtained p-value is
5.2\%, which is consistent with the assumption.

Table~\ref{syserr} shows a summary of the systematic errors.  The dominant
systematic uncertainty in the efficiency of both decay modes in the $P_{tot}
<$100~MeV/{\it{c}} region comes from uncertainty in the Fermi momentum.  In
the proton decay MC, the momentum distribution of the bound nucleon is
simulated based on experimental data~\cite{Nakamura:1976mb}, which is
compared with the Fermi gas model to estimate the uncertainty.  Uncertainties
in energy scale, uniformity of the detector, particle identification, vertex
shift, opening angle, and ring counting are all taken into account as
reconstruction error.  Uncertainties in the 100$\le P_{tot}
<$250~MeV/{\it{c}} region are generally larger because they are mostly bound
protons that suffer from uncertainties associated with $\pi$-FSI, correlated
decays with other nucleons~\cite{Yamazaki:1999gz}, and Fermi motion.  The
leading uncertainties in the background estimate come from cross section,
$\pi$-FSI, and reconstruction. An additional systematic error for neutron
tagging is assigned only for SK-IV. As shown in Figure~\ref{neutron} and Ref.~\cite{Irvine:2014hja}, neutron multiplicity in atmospheric neutrino data and MC agree well. Also,
 as described in Section III, the data with neutron source is compared with MC
in several detector positions. The systematic error on neutron tagging is assigned to be
10\% estimated by the maximum deviation between data and MC.
The run time of SK is well defined and systematic
uncertainty in the exposure is negligibly small (1\%).
        
\begin{figure}[htbp]
  \begin{center}
   \includegraphics[width=80mm]{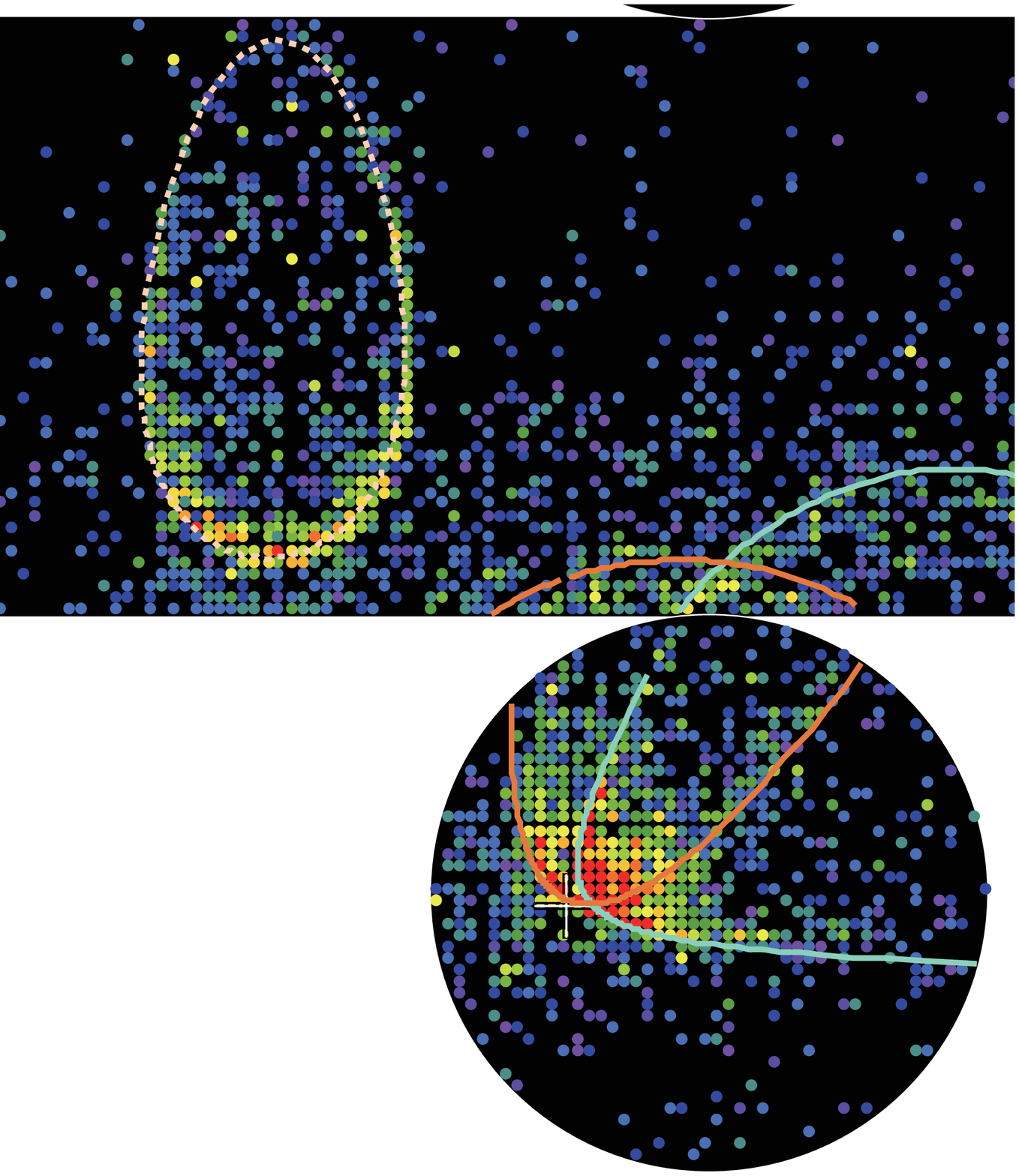}
  \end{center}
 \caption{ \protect \small (color online) Event display of the second
   candidate event, zoomed to the region of the rings.  The blue solid line
   and the tan dashed line show the reconstructed $e$-like and $\mu$-like
   ring, respectively. The dark orange solid line shows an additional $e$-like
   ring that was identified in the initial ring counting process, but it is
   rejected by the ring correction because it is too close in angle to the
   other $e$-like ring (blue line). As a result, this event is judged as a
   two-ring event.  }
  \label{cand2}
\end{figure}

Both of the observed candidates for $p \rightarrow \mu^{+} \pi^{0}$ are
located near the boundary of the signal region.  The first candidate has a
reconstructed proton mass of 903$\pm$19~MeV/$c^2$ and 248$\pm$5~MeV/{\it{c}}
total momentum.  The first candidate is categorized as a two-ring event; the
$e$-like ring has momentum 375~MeV/{\it{c}}, the $\mu$-like ring has
551~MeV/{\it{c}} momentum, and there is a 158~degree opening angle between
the two rings.  For the second candidate, the ring counting algorithm
originally found one $\mu$-like ring and two $e$-like rings. However, one of
$e$-like rings (orange ring in Figure~\ref{cand2}) is judged as a fake ring
at the final stage of ring counting. This final stage discards rings caused
by multiple coulomb scattering of charged particles, which is done by
examination of a ring's angle relative to other rings and by visible energy;
it is applied to both data and MC.  As a result, the second candidate is
judged as a two-ring event.  In the final result of the reconstruction, the
second candidate has a reconstructed proton mass of 832$\pm$17~MeV/$c^2$ and
a total momentum of 238$\pm$5~MeV/{\it{c}} calculated from the remaining
$e$-like ring with 461~MeV/{\it{c}}, the $\mu$-like ring with
391~MeV/{\it{c}}, and the 149~degree angle between the two rings.  If the
third ring had not been rejected, the reconstructed invariant mass from the
two $e$-like rings would have been 104~MeV/$c^2$, which could be gammas from
$\pi^0$ decay.  The total momentum and proton mass would increase to
289~MeV/{\it{c}} and 880~MeV/$c^2$, respectively, which would move the event
outside of the defined signal region
\footnote{ This event is judged as an 3 ring event, but leaves the
  signal box if we use updated PMT gain correction, introduced in 2016, which
  depends on PMT production year.}.

\begin{table*}
  \begin{center}
    \begin{tabular}{ll|cccc}
      \hline \hline
                    &                 &  SK-I       &  SK-II      &  SK-III     & SK-IV      \\
      \hline 
Exp. & kt$\cdot$yrs                 &  91.7       &  49.2       &  31.9       & 133.5       \\
\hline
$p \rightarrow e^{+} \pi^{0}$    &             &  &  &  & \\
Low $P_{tot}$    & Eff.(\%)  & 18.8$\pm$1.9 & 18.3$\pm$1.9 & 19.6$\pm$2.0 & 18.7$\pm$1.9 \\
                     & BKG  & $0.03^{+0.03}_{-0.02}$ & $<$0.01 & $<$0.01 & $0.02^{+0.03}_{-0.02}$ \\
                     & (/Mt$\cdot$yr)  & $0.36^{+0.30}_{-0.20}$ & $0.26^{+0.27}_{-0.17}$ & $0.09^{+0.21}_{-0.08}$ & $0.18^{+0.25}_{-0.13}$ \\                     
                     & OBS  & 0 & 0 & 0& 0 \\ 
High $P_{tot}$    & Eff.(\%)  & 20.4$\pm$3.6 & 20.2$\pm$3.6 & 20.5$\pm$3.6 & 19.4$\pm$3.4 \\
                     & BKG  & 0.22$\pm$0.08 & 0.12$\pm$0.04 & 0.06$\pm$0.02& 0.15$\pm$0.06 \\
                     & (/Mt$\cdot$yr)  & 2.4$\pm$0.8 & 2.5$\pm$0.9 & 1.8$\pm$0.7& 1.1$\pm$0.3 \\                     
                     & OBS  & 0 & 0 & 0& 0 \\
\hline   
$p \rightarrow \mu^{+} \pi^{0}$    &             &  &  &  & \\
Low $P_{tot}$    & Eff.(\%)  & 16.4$\pm$1.5 & 16.0$\pm$1.5 & 16.4$\pm$1.5 & 20.1$\pm$1.9 \\
                     & BKG  & $0.03^{+0.02}_{-0.02}$ & $<$0.01 & $<$0.01 & $0.01^{+0.02}_{-0.01}$ \\
                     & (/Mt$\cdot$yr)  & $0.31^{+0.26}_{-0.17}$ & $0.10^{+0.13}_{-0.07}$ & $0.22^{+0.22}_{-0.14}$ & $0.09^{+0.21}_{-0.08}$ \\                     
                     & OBS  & 0 & 0 & 0& 0 \\ 
High $P_{tot}$    & Eff.(\%)  & 15.3$\pm$2.8 & 15.3$\pm$2.8 & 16.5$\pm$3.0 & 18.2$\pm$3.3 \\
                     & BKG  & 0.33$\pm$0.10 & 0.14$\pm$0.05 & 0.12$\pm$0.04 & 0.23$\pm$0.08 \\
                     & (/Mt$\cdot$yr)  & 3.6$\pm$1.1 & 2.9$\pm$0.9 & 3.7$\pm$1.2& 1.7$\pm$0.6 \\                      
                     & OBS  & 0 & 0 & 0 & 2 \\                     
      \hline \hline
    \end{tabular}
  \caption{\protect \small Summary of the $p \rightarrow e^{+} \pi^{0}$ and
    $p \rightarrow \mu^{+} \pi^{0}$ proton decay modes. The selection
    efficiencies and expected backgrounds with quadratic sum of statistical
    and systematic errors, and number of observed events are shown for each
    detector period. Low $P_{tot}$ and high $P_{tot}$ are defined as $P_{tot}
    <$~100MeV/{\it{c}} and 100$\le P_{tot} <$~250MeV/{\it{c}}, respectively.
  }
  \label{result}
  \end{center}
\end{table*}

\begin{table*}
  \begin{center}
    \begin{tabular}{ll|cc|cc}
      \hline \hline
                    &                     & \multicolumn{2}{c|}{$p \rightarrow e^{+} \pi^{0}$} & \multicolumn{2}{c}{$p \rightarrow \mu^{+} \pi^{0}$} \\
      \hline 
    &                     &  low $P_{tot}$       &  high $P_{tot}$       &  low $P_{tot}$       & high $P_{tot}$       \\
\hline
Efficiency    &             &  &  &  & \\
    & $\pi$-FSI  & 2.8 & 10.6 & 2.9 & 12.1 \\
    & Corr. decay & 1.9 & 9.1 & 1.7 & 9.0 \\
    & Fermi mom.  & 8.5 & 9.3 & 8.0 & 9.6 \\ 
    & Reconstruction  & 4.6 & 5.6 & 3.7 & 3.3 \\
\hline
    & Total  & 10.2 & 17.7 & 9.4 & 18.2 \\
\hline \hline   
Background    &             &  &  &  & \\
       & Flux  & 7.0 & 6.9 & 7.0 & 7.0 \\
       & Cross section & 14.5 & 10.4 & 8.4 & 7.8 \\
       & $\pi$-FSI  & 15.4 & 15.4 & 14.2& 14.4 \\ 
       & Reconstruction & 21.7 & 21.7 & 21.7 & 21.7 \\
       & (neutron tag)  & 10 & 10 & 10 & 10 \\
\hline
       & Total (I/II/III) & 31.2 & 29.4 & 28.1 & 28.1 \\  
       & (IV)  & 32.7 & 31.1 & 29.9 & 29.8 \\ 
\hline \hline   
Exposure &             &1.0  &1.0  &1.0  &1.0 \\ 
      \hline \hline
    \end{tabular}
  \caption{\protect \small Summary of systematic errors for efficiencies,
    backgrounds, and exposures in percent (\%). Low $P_{tot}$ and high
    $P_{tot}$ are defined as $P_{tot} <$~100MeV/{\it{c}} and 100$\le P_{tot}
    <$~250MeV/{\it{c}}, respectively. For SK-IV only, a neutron tagging
    uncertainty is included, and the total systematic error for SK-IV is also
    shown including this uncertainty, both in parentheses.  }
  \label{syserr}
  \end{center}
\end{table*}

%\begin{figure}[htbp]
%  \begin{center}
%   \includegraphics[width=90mm]{cand1-normal-view3.eps}
%  \end{center}
% \caption{ \protect \small (color online) Event display of the first
%   observed event. The blue solid line and the yellow dashed line show
%   the reconstructed $e$-like and $\mu$-like ring, respectively.  }
%  \label{cand1}
%\end{figure}

\section{LIFETIME LIMIT} 
The observed events are consistent with expected backgrounds and a proton
lifetime limit is calculated using a Bayesian method~\cite{Amsler:2008zzb}
\cite{Roe:2000fe}.  In Ref.~\cite{Abe:2014mwa}, the proton lifetime was
calculated combining different search methods as well as different run
periods. The same method is used here for combining $P_{tot}
<$~100MeV/{\it{c}} and 100$\le P_{tot} <$~250 MeV/{\it{c}}.  For measurements
$i=$ $1-4$ ($P_{tot} <$~100MeV/{\it{c}}, SK-I--SK-IV) and $5-8$ (100~$\le
P_{tot} <$~250 MeV/{\it{c}}, SK-I--SK-IV), the conditional probability
distribution for the decay rate is expressed as:
\begin{eqnarray}
P(\Gamma|n_i) &=& \iiint \frac{e^{-(\Gamma\lambda_i\epsilon_i+b_i)}
(\Gamma\lambda_i\epsilon_i+b_i)^{n_i}}{n_i!} \times \nonumber \\ \nopagebreak
 & & ~~P(\Gamma)P(\lambda_i)P(\epsilon_i)P(b_i)
     \,d\lambda_i\,d\epsilon_i\,db_i
\end{eqnarray}
where $n_{i}$ is the number of candidate events in the $i$-th proton decay
search, $\lambda_i$ is the true detector exposure, $\epsilon_i$ is the true
detection efficiency, and $b_i$ is the true number of background events. The
decay rate prior probability distribution $P(\Gamma)$ is 1 for $\Gamma \ge 0$
and otherwise 0. $P(\lambda_i)$, $P(\epsilon_i)$, and $P(b_i)$ are the prior
probability distributions for detector exposure, efficiency, and background,
respectively, which are assumed to be Gaussian distributions with $\sigma$
described in Table~\ref{syserr}~\cite{Abe:2014mwa}.
 The lower limit of the nucleon decay rate,
$\Gamma_{\mathrm{limit,}}$ is:

\begin{eqnarray}
\mathrm{CL}=\frac{\int^{\Gamma_{\mathrm{limit}}}_{\Gamma=0}\prod^{N=8}_{i=1}P(\Gamma|n_i)\,d\Gamma} {\int^\infty_{\Gamma=0}\prod^{N=8}_{i=1}P(\Gamma|n_i)\,d\Gamma},
\end{eqnarray}
\noindent where CL is the confidence level,
  taken to be 90\%. The lower lifetime limit of $p \rightarrow l^{+} \pi^{0}$ 
($l^{+}$ denotes $e^{+}$ or $\mu^{+}$) is given by:
\begin{eqnarray}
\tau/\mathrm{B}_{p\rightarrow l^{+} \pi^0}=\frac{1}{\Gamma_{\mathrm{limit}}}.
\end{eqnarray}

The results of the limit calculation combining the two regions are:
\begin{eqnarray}
\tau/\mathrm{B}_{p\rightarrow e^{+} \pi^0} > 
  1.6 \times 10^{34} \mathrm{years,} \nonumber
\end{eqnarray}
\begin{eqnarray}
\tau/\mathrm{B}_{p\rightarrow \mu^{+} \pi^0} > 
  7.7 \times 10^{33} \mathrm{years,} \nonumber
\end{eqnarray}
at the 90\% confidence level.  $\tau/\mathrm{B}_{p\rightarrow \mu^{+} \pi^0}$
is lower than $\tau/\mathrm{B}_{p\rightarrow e^{+} \pi^0}$, and this is also
lower than our previous publication~\cite{Nishino:2012bnw} because of the two
observed events, which are consistent with atmospheric neutrino background.

\section{CONCLUSION} 
We analyzed 0.306 megaton$\cdot$years of Super-Kamiokande data to search for
proton decay via $p \rightarrow e^{+} \pi^{0}$ and $p \rightarrow \mu^{+}
\pi^{0}$. Neutron tagging was introduced in SK-IV and it succeeds in
rejecting half of the backgrounds. The signal region from SK-I to SK-IV was
divided into two regions of $P_{tot}$ to obtain better sensitivity.  We
observed 0 events (0.07 and 0.54 expected background in lower and higher 
momentum box, respectively)  for $p \rightarrow e^{+} \pi^{0}$
and 2 events in the higher momentum box (0.05, and 0.82 expected background 
in lower and higher momentum box, respectively) for $p \rightarrow \mu^{+}
\pi^{0}$. The obtained proton lifetime limits at 90\% confidence level are
$>1.6 \times 10^{34}$ years for $p \rightarrow e^{+} \pi^{0}$ and $>7.7
\times 10^{33}$ years for $p \rightarrow \mu^{+} \pi^{0}$. \\

\section{ACKNOWLEDGMENTS} 
We gratefully acknowledge the cooperation of the Kamioka Mining and Smelting Company. The Super-Kamiokande experiment has been built and operated from funding by the Japanese Ministry of Education, Culture, Sports, Science and Technology, the U.S. Department of Energy, and the U.S. National Science Foundation. Some of us have been supported by funds from the Research Foundation of Korea (BK21 and KNRC), the Korean Ministry of Science and Technology, the National Research Foundation of Korea (NRF-20110024009), the European Union H2020 RISE-GA641540-SKPLUS), the Japan Society for the Promotion of Science, the National Natural Science Foundation of China under Grants No. 11235006, the National Science and Engineering Research Council (NSERC) of Canada, and the Scinet and Westgrid consortia of Compute Canada.
 
% Create the reference section using BibTeX:
%\bibliography{basename of .bib file}
\bibliography{reference}

\end{document}